\documentclass[%
 aip,
 jmp,%
 amsmath,amssymb,
preprint,%
]{revtex4-1}

\usepackage{graphicx}
\usepackage{dcolumn}
\usepackage{physics}
\usepackage{bm}
\usepackage[dvipsnames]{xcolor}
\usepackage{soul}

\begin{document}


\title[]{Spin relaxation due to the D'yakonov-Perel' mechanism in 2D semiconductors with an elliptic band structure}

\author{Seyed M. Farzaneh}
 \email{farzaneh@nyu.edu}
 \altaffiliation[]{}
\author{Shaloo Rakheja}%
 \email{shaloo.rakheja@nyu.edu}
\affiliation{ 
Department of Electrical and Computer Engineering, New York University
}%

\date{November 22, 2018}

\begin{abstract}
D'yakonov--Perel' (DP) mechanism describes the dynamics of non--equilibrium spin distribution in a two--dimensional (2D) system in the presence of Rashba spin--orbit coupling. In this paper, we study the anisotropy of spin relaxation via the DP mechanism for a 2D semiconductor with an elliptic band structure.
Within the effective--mass approximation, the low--energy band structure is described using anisotropic in--plane effective mass of free carriers.
Spin relaxation time of free carriers is calculated theoretically using the time evolution equation of the density matrix of a polarized spin ensemble in the strong momentum scattering regime.
Results are obtained for scattering potential due to both Coulomb interaction and neutral defects in the sample.
We show that the ratio of spin relaxation time in the y-- and x--direction within the 2D plane displays a power--law dependence on the effective mass ratio, while the exponent captures the details of the scattering potential.
The model is applied to study electron spin relaxation in monolayer black phosphorus, which is known to exhibit significant band structure ellipticity.
The model can also predict spin relaxation anisotropy in mechanically strained 2D materials in which elliptic band structure emerges as a consequence of the modification of the lattice constants.\\
\end{abstract}

\maketitle

\section{introduction}
The study of non--equilibrium spin relaxation in metals and semiconductors is key in analyzing experimental data and enabling spin-based device applications.
For most spin-based technologies, a long spin relaxation time is desirable. However, fast switching may be achieved in certain devices if the spin relaxation time is sufficiently short.\cite{nishikawa1995all, hall1999subpicosecond}
As such, material systems that allow for spin relaxation engineering are important from a practical standpoint.
Two-dimensional (2D) materials, such as 
semimetallic graphene and semiconducting black phosphorus (BP), have a weak intrinsic spin--orbit interaction and are expected to have long spin relaxation times, which could allow spin--encoded information to travel macroscopic distances in these materials.
Moreover, an external electric field perpendicular to the plane of the 2D materials is an effective method to tune their spin--transport characteristics.\cite{min2006intrinsic, han2011spin, avsar2017gate, kurpas2018spin}
The literature on the dynamics and transport of nonequilibrium spin in graphene is rich and well--established as demonstrated in a number of experiments on graphene--based spin--valve devices.\cite{hill2006graphene, tombros2007electronic, han2011spin, ingla201524, drogeler2016spin}
Recent experiments have revealed spin relaxation times on the order of
a few nanoseconds in graphene.\cite{ingla201524, drogeler2016spin}
A similar value of spin relaxation time (few nanoseconds) has also been measured in ultra-thin BP in a recent experiment.\cite{avsar2017gate}
Unlike graphene, BP has a highly anisotropic band structure with an elliptic Fermi contour due to its puckered honeycomb structure.
Additionally, the effect of mechanical strain on the band structure anisotropy is more pronounced in the case of BP.\cite{peng2014strain} 
The anisotropy in the band structure leads to anisotropic momentum relaxation,\cite{ong2014anisotropic, liu2016mobility, liu2017temperature} anisotropic Rashba spin-orbit coupling,\cite{popovic2015electronic} and, therefore, anisotropic spin relaxation.\cite{kurpas2016spin}

The goal of this work is to theoretically investigate spin relaxation in intrinsically anisotropic BP as well as other 2D materials in which mechanical strain results in ellipticity of the band structure. 
Our focus is only on the time--domain dynamics of non--equilibrium homogeneous spin distribution in 2D semiconductors; hence, diffusion terms in spin Boltzmann kinetic equations are not included. 
Within the D'yakonov-Perel' (DP) theory, spin relaxation occurs due to the scattering--induced motional narrowing of spin precession about the Rashba spin--orbit field which is the result of broken inversion symmetry in the presence of an external electric field.\cite{dyakonov1972spin, bychkov1984properties}
In the motional narrowing regime, stronger momentum scattering leads to longer spin lifetimes.
In centro--symmetric structures like BP, the Elliott-Yafet (EY) mechanism also leads to spin relaxation.
Per the EY mechanism, spins flip at momentum scattering events which eventually cause nonequilibrium spin population to relax.\cite{elliott1954theory}
Recent experimental studies suggest that the spin relaxation in BP is dominated by the EY mechanism.\cite{avsar2017gate, li2014electrons} 
However, in the presence of external electric fields or with the use of polar substrates required for the deposition of BP thin films,\cite{popovic2015electronic} it is important to consider the DP mechanism while evaluating the spin relaxation time in BP. 
This work focuses on spin dynamics processes in BP monolayers and other anisotropic 2D materials in the presence of symmetry breaking electric fields for which the DP mechanism is relevant.  
The main result in this paper is that the ratio of in--plane components of spin relaxation time ($\tau_{s,yy}/\tau_{s,xx}$) due to the DP mechanism 
scales as $(m_y/m_x)^\nu$ where $\nu$ is a function of the scattering potential. Here, $m_y$ and $m_x$ denote the effective carrier mass in $y$-- and $x$--direction, respectively.
For Coulomb scatterers $\nu\simeq0.5$ whereas for neutral defects $\nu\simeq0.5-1.5$ depending on the range of the potential.

Previous studies present a closed--form solution of the spin relaxation time due to the DP mechanism in isotropic 2D semiconductors.\cite{fabian2007semiconductor, averkiev2002spin}
Recently, DP relaxation mechanism in BP was studied using this closed--form solution .\cite{kurpas2016spin, kurpas2018spin}
In this closed--form solution, the momentum scattering time is introduced as a constant. 
In our work, we replace the closed--form solution with an implicit one that captures the anisotropy of momentum scattering in the presence of Coulomb and neutral defect scatterers. 
That is because the electrons experience different scattering rates in different directions and the spin relaxation rates should be evaluated accordingly.
To do so, we consider a general elliptic band structure within the effective--mass approximation. 
Representing spin polarization with density matrices, we study the spin dynamics via the time evolution equation of the density matrix, which includes spin precession and momentum scattering processes. 
Finally, solving the time evolution equation in the quasi--static regime where momentum relaxation is much faster than spin relaxation, we calculate the relaxation rate of the total spin polarization. 
Calculations are performed in the low temperature regime where only the Fermi level is taken into account. 
High-temperature effects are not expected to change the anisotropy of spin relaxation. The only effect will be on the absolute values of the spin relaxation time.\cite{avsar2017gate}  
Also, we ignore any magnetic order emerging at low temperatures.\cite{seixas2015atomically}

\section{Theoretical Model}
An external electric field applied perpendicularly, along the $\vu*{z}$ axis, to a two-dimensional semiconductor breaks the inversion symmetry and, therefore, introduces the Rashba spin--orbit Hamiltonian
\begin{equation}
    H_\text{R} = \frac{\lambda_\text{R}}{2}(\vb*{p}\times\vu*{z})\vdot\vb*{\sigma}\,,
\end{equation}
where $\vb*{\sigma}$ is the vector of Pauli matrices and $\lambda_\text{R}$ is the strength of spin--orbit coupling which is proportional to the magnitude of the electric field and the proportionality constant depends on the details of crystal structure. 
For the conduction band of a generic anisotropic semiconductor consisting of light atoms (such as black phosphorus), it can be shown (Appendix), using $\vb*{k}\vdot\vb*{p}$ perturbation theory, that the $\vb*{k}$--space Hamiltonian is $H=H_0 + H_{\vb*{k}}$ where
\begin{subequations}
    \begin{equation}
        H_0 = E_\text{c} + \frac{\hbar^2k_x^2}{2m_x} + \frac{\hbar^2k_y^2}{2m_y}\,,
    \end{equation}
    \begin{equation}
        H_{\vb*{k}} = \frac{\lambda_\text{R}}{2}(\vb*{\tilde{k}}\times\vu*{z})\vdot\vb*{\sigma}\cdot
    \end{equation}
\end{subequations}
The conduction band minimum is denoted by $E_\text{c}$, $m_x$ and $m_y$ are the effective masses along the $\vu*{x}$ and $\vu*{y}$ axes, $\vb*{\tilde{k}}=(m_0/m_x)k_x\vu*{x} + (m_0/m_y)k_y\vu*{y}$, $m_0$ is the free electron mass, and the Rashba term $H_{\vb*{k}}$ is considered as a small perturbation compared to $H_0$.
We note that the anisotropic Rashba Hamiltonian $H_{\vb*{k}}$ is similar to the C$_{2v}$ symmetry preserving Hamiltonian for monolayer BP proposed in Ref. \onlinecite{kurpas2018spin} which is given as $H_\text{SOC}=(\alpha + \beta)k_y\sigma_x + (\alpha - \beta)k_x\sigma_y$. The two Hamiltonians are connected by choosing $\alpha=(m_0/m_x + m_0/m_y)\lambda_\text{R}/4$ and $\beta=(m_0/m_x - m_0/m_y)\lambda_\text{R}/4$.
For energies $E>E_\text{c}$, the Fermi contour is an ellipse described by $k_x^2/a^2 + k_y^2/b^2 = 1$ where $a=\sqrt{2m_x\qty(E - E_\text{c})}/\hbar$ and $b=\sqrt{2m_y\qty(E - E_\text{c})}/\hbar$.
The Fermi contour can also be described in polar coordinates where the magnitude of the in--plane wavevector is a function of the polar angle $\theta$, i.e.  $k\qty(\theta)=ab/\sqrt{a^2\sin^2\theta + b^2\cos^2\theta}$.
We can rewrite $H_{\vb*{k}}$ as $\frac{\hbar}{2}\vb*{\Omega}_{\vb*{k}}\vdot\vb*{\sigma}$ where $\vb*{\Omega}_{\vb*{k}}$ is the effective Rashba spin--orbit field given as
\begin{equation}
\label{eq:field}
    \vb*{\Omega}_{\vb*{k}} = \lambda_\text{R}\qty( \frac{m}{m_y}k\qty(\theta)\sin\theta\vu*{x} -\frac{m}{m_x}k\qty(\theta)\cos\theta\vu*{y})\cdot
\end{equation}
To the conduction electrons, $\vb*{\Omega}_{\vb*{k}}$ acts as a $\vb*{k}$--dependent magnetic field. Electrons with different momenta precess around different axes. 
Therefore, scattering between different momenta randomizes the precession of a polarized ensemble and consequently leads to spin relaxation. 
This is the aforementioned DP relaxation mechanism. 
To calculate the spin relaxation time due to the DP mechanism, we follow a similar procedure as in Refs. \onlinecite{averkiev2002spin, fabian2007semiconductor}, but we specifically analyze 2D materials with an elliptic band structure.
A spin--polarized ensemble described by a $\vb*{k}$--dependent density matrix $\rho_{\vb*{k}}$ is considered. The time evolution of a spin ensemble in the absence of inhomogeneities and spin drift due to external fields is~\cite{averkiev2002spin}
\begin{equation}
    \label{eq:evolution}
    \pdv{\rho_{\vb*{k}}}{t} = \frac{i}{\hbar}[\rho_{\vb*{k}}, H_{\vb*{ k}}] -
    \sum_{\vb*{k'}\not={\vb* k}}W_{\vb*{kk'}}(\rho_{\vb*{k}} - \rho_{\vb*{ k'}})\,,
\end{equation}
where $W_{\vb*{kk'}}$ is the probability density of transition between $\vb*{k}$ and $\vb*{k'}$ states. The first term on the right-hand side represents spin precession about the effective Rashba field, and the second term represents momentum scattering between incoming wavevector $\vb*{k}$ and outgoing wavevector $\vb*{k'}$. We can decompose the density matrix as $\rho_{\vb*{k}}=\overline{\rho} + \rho'_{\vb*{ k}}$, where $\overline{\rho}$ is the average of density matrix over different ${\vb*{k}}$'s of the Fermi contour and $\rho'_{\vb*{k}}$ is a small perturbation with zero average, i.e. $\overline{\rho'_{\vb*{k}}}=0$. 
Taking the average of Eq. \ref{eq:evolution} over the Fermi contour, we obtain
\begin{equation}
    \label{eq:evolution-1}
    \pdv{\overline{\rho}}{t} = 
    \frac{i}{\hbar}\overline{[\rho'_{\vb*{k}}, H_{\vb*{k}}]}\,,
\end{equation}
where we used the fact that $\overline{H_{\vb*{k}}}$ is zero. The reason is that for each point $\vb*{k}$ on the Fermi contour, $-\vb*{k}$ is also on the Fermi contour. Since $H_{\vb*{k}}$ is linear in $\vb*{k}$ and therefore an odd function of $\vb*{k}$, i.e. ${H_{\vb*{-k}}}=-H_{\vb*{k}}$, its average over the Fermi contour is zero.
Applying the decomposition to Eq. \ref{eq:evolution} and dropping the terms containing product of $H_{\vb*{k}}$ and $\rho'_{\vb*{k}}$, we can find the quasistatic value of $\rho'_{\vb*{k}}$, by setting $\pdv*{\rho'_{\vb*{k}}}{t}$ to zero, assuming that the momentum relaxation is much faster than spin relaxation.
Therefore, 
\begin{equation}
    \label{eq:evolution-2}
    \frac{i}{\hbar}[\overline{\rho}, H_{\vb*{k}}] =
    \sum_{\vb*{k'}\not=\vb*{k}}W_{\vb*{kk'}}(\rho'_{\vb*{k}} - \rho'_{\vb* {k'}})\,\cdot
\end{equation}
Equations \ref{eq:evolution-1} and \ref{eq:evolution-2} are coupled and must be solved self-consistently. 
To do so, first we assume that the average spin polarization is in $\vu*{s}$ direction. 
Therefore, we can write $\overline{\rho} = \frac{1}{2} + \vu*{s}\vdot\vb*{\sigma}$.
It can be shown that $\frac{i}{\hbar}[\overline{\rho}, H_{\vb*{k}}] = -(\vu*{ s}\times\vb*{\Omega}_{\vb*{k}})\vdot{\vb* \sigma}$.
Using Eq. \ref{eq:evolution-2}, we can solve for $\rho'_{\vb*{k}}$ iteratively using the following equation:
\begin{equation}
    \label{eq:rho_p}
    \rho'_{\vb*{k}} = \frac{-(\vu*{s}\times\vb*{\Omega}_{\vb*{k}})\vdot{\vb* \sigma} + \sum_{\vb*{k'}\not=\vb*{k}}W_{\vb*{kk'}}\rho'_{\vb*{k'}}}{\sum_{\vb*{k'}\not=\vb*{k}}W_{\vb*{kk'}}}\,\cdot
\end{equation}
Plugging $\rho'_{\vb*{k}}$ into Eq. \ref{eq:evolution-1}, we can calculate the rate of decay $\pdv* {\overline{\rho}}{t}$ or correspondingly $\dv*{\vu*{s}}{t}=-\vu*{s}/\tau_s$ which results in the spin relaxation time $\tau_s$.

The collision sum in the continuum limit becomes an integral, i.e. $\sum_{\vb*{k'}\not=\vb*{k}}W_{\vb*{kk'}}\rightarrow A\int d^2\vb*{k}(2\pi)^{-2}W_{\vb*{kk'}}$, where $A$ is the area of the 2D semiconductor. 
Using Fermi's golden rule, the transition rate is given as $W_{\vb*{kk'}}=\frac{2\pi}{\hbar}N\abs{\mel{\vb*{k}}{V}{\vb*{k'}}}^2\delta(E({\vb* k}) - E({\vb*{k'}}))$, where $N$ is the number of scatterers and $\mel{{\vb* k}}{V}{\vb*{k'}}$ is the matrix element of the scattering potential $V$ given as
\begin{equation}
    \label{eq:elements}
    \mel{{\vb* k}}{V}{\vb*{k'}} = \int d^2\vb*{r} \psi^*_{\vb*{k}}\qty({\vb* r})V\qty({\vb* r})\psi_{\vb*{k'}}({\vb* r})\cdot
\end{equation}
Replacing $\psi_{\vb*{k}}({\vb* r})$ with Bloch wave functions $A^{-1/2}\text{e}^{i{\vb* k}\vdot{\vb* r}}u_{\vb* k}({\vb* r})$, it can be shown that
\begin{equation}
    \mel{{\vb* k}}{V}{\vb*{k'}} = \frac{1}{A}V({\vb* q}) = \frac{1}{A}\int d^2\vb*{r} \text{e}^{-i{\vb* q}\vdot{\vb* r}}V({\vb* r}),
\end{equation}
where ${\vb* q} = \vb*{k} - \vb*{k'}$, $V({\vb* q})$ is the Fourier transform of the scattering potential.
The Coulomb potential is given as $V(\vb*{r})=e^2/4\pi\epsilon_0\epsilon_\text{r}\sqrt{r^2 + d^2}$ where 
$d$ is the depth of the scattering center in the substrate\cite{ando1982electronic}, $\epsilon_0$ is the permittivity of free space, and $\epsilon_\text{r}$ is the relative permittivity of the substrate.
The Fourier transform of $V(\vb*{r})$ is $V({\vb* q})=2\pi \text{e}^2 \exp(-qd)/4\pi\varepsilon_0\varepsilon_\text{r} q$.
For neutral defects $V(\vb*{r})=V_0e^{-r^2/2\sigma^2}$ we obtain $V({\vb* q})=2\pi\sigma^2V_0\exp(-2\pi^2\sigma^2q^2)$, where $V_0$ is the amplitude of the defect potential, and $\sigma$ is the effective potential radius.\cite{yuan2015transport}
The delta function in $W_{\vb*{kk'}}$ reduces the $k$--space integral to an integral over the Fermi contour. Therefore,
\begin{equation}
    \label{eq:momentum-relaxation-final}
    \sum_{\vb*{k'}\not=\vb*{k}}W_{\vb*{kk'}}\rightarrow
    \frac{n}{2\pi\hbar}
    \oint
    d\theta'
    \abs{\pdv{\vb*{k'}}{\theta'}}
    \frac{\abs{V\qty({\vb* q})}^2}{\abs{\grad E\qty(\vb*{k'})}}\,,
\end{equation}
where $n=N/A$ is the density of scatterers and $\abs{\pdv*{\vb*{k'}}{\theta'}}=\sqrt{k^2(\theta) + (dk(\theta)/d\theta)^2}$. 
Finally, the average over the Fermi contour in Eq. \ref{eq:evolution-1} for a given function $f({\vb* k})$ is defined as $\overline{f} = \ell^{-1}\oint d\theta\abs{\dv*{\vb*{k}}{\theta}}f({\vb* k})$,
where $\ell$ is the perimeter of the Fermi contour.

\section{results}
\begin{figure*}
        \includegraphics[scale=1.0]{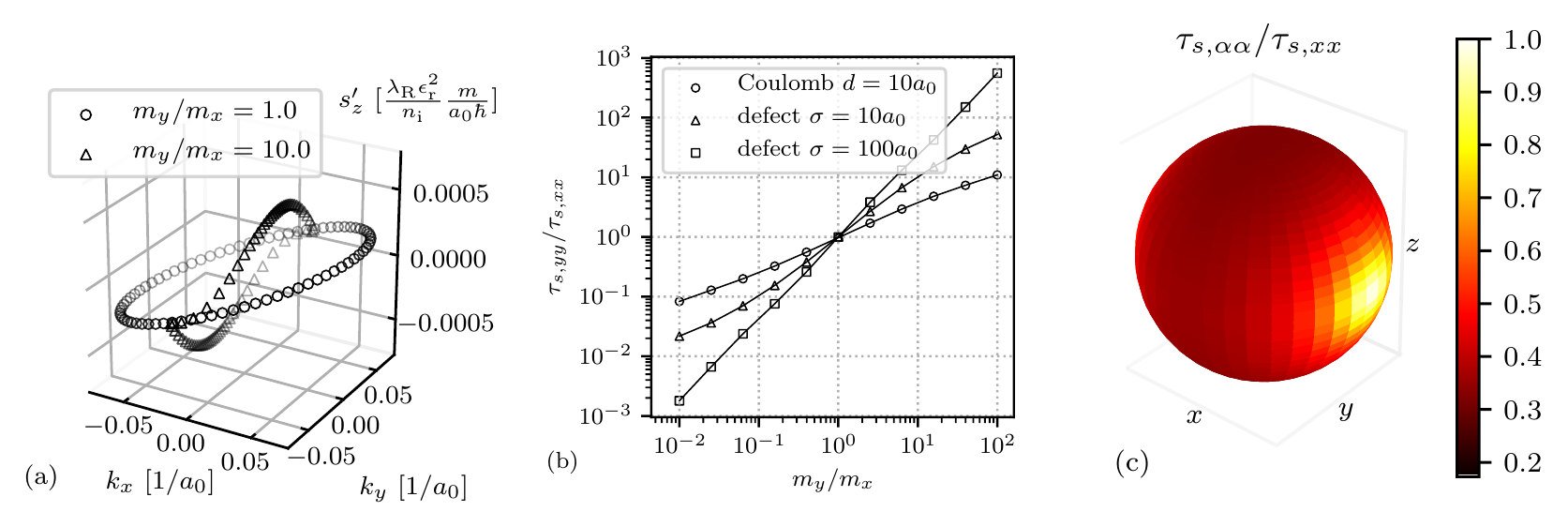} 
    \vspace{-0.1in}
    \caption[example] 
    {(a) Spin perturbation $s'_z = \Tr(\rho'_{\vb* k}\sigma_z)$ evaluated using Eq. \ref{eq:rho_p} for a spin ensemble initially polarized along the $\vu*{x}$ axis in the presence of Coulomb potential.  
    Here $m_x$ and $m_y$ are in--plane effective masses along the $\vu*{x}$ and $\vu*{y}$ axes respectively, $\lambda_\text{R}$ is the Rashba strength, $n_\text{i}$ is the density of Coulomb scatterers, and $\epsilon_\text{r}$ is the effective permittivity of the substrate. 
    (b) In-plane anisotropy of spin relaxation time, $\tau_{s,yy}/\tau_{s,xx}$, versus effective mass ratio. Here the parameters of the scattering potentials are $d=10a_0$, $V_0=10$ eV, $\epsilon_\text{r}=3.8$, and $\sigma=10a_0$ and $\sigma=100a_0$ for short-- and long--range defects, respectively.
    (c) Normalized spin relaxation time as a function of initial polarization direction.
    Also $a_0$ is the Bohr radius and $E-E_\text{c} = 0.1$ eV for all plots.
    }
    \label{fig:anisotropy} 
\vspace{-0.1in}
\end{figure*}

The behavior of the perturbation density matrix $\rho'_{\vb* k}$ is examined by calculating the corresponding perturbation in spin. 
We assume that the initial spin polarization is along the $\vu*{x}$ axis by replacing $\vu*{s}$ with $\vu*{x}$ in Eq. \ref{eq:rho_p}.
Therefore, $\rho'_{\vb* k}$ contains only the $\sigma_z$ component, and the spin perturbation exists only along the $\vu*{z}$ axis. 
That is $s'_z=\Tr(\rho'_{\vb* k}\sigma_z)$.
Figure \ref{fig:anisotropy}a illustrates $s'_z$ over the Fermi contour in the presence of Coulomb potential for two values of anisotropy, i.e. the effective mass ratio $m_y/m_x$. 
The isotropic $\rho'_{\vb* k}$, regardless of the scattering potential, is described by the closed--form solution to Eq. \ref{eq:evolution-2} that is  $\frac{i\tau}{\hbar}[\overline{\rho}, H_{\vb*{k}}]$ where $\tau$ is a time constant closely related to the momentum relaxation time $\tau_p$.\cite{fabian2007semiconductor}
As evident from Fig. \ref{fig:anisotropy}a, the anisotropic curve is very different from the isotropic curve and cannot be described simply by the closed--form solution. Hence, a direct evaluation of Eq.~\ref{eq:rho_p} becomes inevitable.
Similarly, in the case of defects, the transition probability $W_{\vb*{kk'}}$ is also $k$--dependent and the anisotropic $s'_z$ will have the shape of a distorted ellipse (not shown in the figure).
We note that $s'_z$ plotted in Fig. \ref{fig:anisotropy}a is in atomic units and proportional to $\lambda_\text{R}\epsilon^2_\text{r}/n_\text{i}$ where $n_\text{i}$ is the density of Coulomb scatterers.  
As long as the momentum scattering is strong enough, i.e. $(\lambda_\text{R}\epsilon^2_\text{r}/n_\text{i})(m/a_0\hbar)\ll1$ or $s'_z\ll 1$, the perturbation is much less than the average polarization, i.e. $\rho'_{\vb* k}\ll \overline{\rho}$, and the assumption in deriving Eq. \ref{eq:evolution-2} remains valid. 
In the case of defects, a similar condition holds, i.e. $(\lambda_\text{R}/n_\text{d})(m/a_0\hbar)\ll1$ where $n_\text{d}$ is the density of defects.

The effect of band structure ellipticity is further examined in Fig.~\ref{fig:anisotropy}b. In this figure, $\tau_{s,\alpha\alpha}$ denotes the spin relaxation time for an ensemble initially polarized in the direction of $\vu*{\alpha}$.
We note that spin relaxation rates of the form $1/\tau_{s,\alpha\beta}$ for $\alpha\bot\beta$ are equal to zero; in other words there is no 
spin dephasing.
As the anisotropy increases, the ratio of in--plane spin relaxation time $\tau_{s,yy}/\tau_{s,xx}$ increases proportional to 
$(m_y/m_x)^\nu$ where $\nu$ is a constant that depends on the details of the scattering potential. 
Our results show that for Coulomb potential $\nu\simeq0.5$ whereas for neutral defects $\nu\simeq0.5-1.5$ depending on the range of the potential $\sigma$.
We also note that the spin relaxation time is longer in the direction of the heavier effective mass.
The spin relaxation for an ensemble polarized along the $\vu*{z}$ axis is always faster than in--plane directions (not shown in the figure). 
Replacing $\vu*{s}$ with $\vu*{z}$ in Eq. \ref{eq:rho_p}, we can see that $\rho'_{\vb*{k}}$ obtains both $\sigma_x$ and $\sigma_y$ components. Therefore, the corresponding spin relaxation rate is the sum of relaxation rates along the in-plane directions, i.e. $1/\tau_{s,zz} = 1/\tau_{s,xx} + 1/\tau_{s,yy}$. It is evident that for the isotropic case, i.e. $m_x=m_y$ we obtain $\tau_{s,xx}=\tau_{s,yy}=2\tau_{s,zz}$ which has been reported previously in the literature.\cite{fabian2007semiconductor} 
Figure \ref{fig:anisotropy}c illustrates the normalized spin relaxation rate $1/\tau_{s,\alpha\alpha}$ as a function of initial polarization direction $\vu*{\alpha}$ for an effective mass anisotropy of $m_y/m_x=0.1$. As expected the spin polarization in the $x$--direction is preserved longer than other directions.

\begin{figure*}
	\includegraphics[scale=1.0]{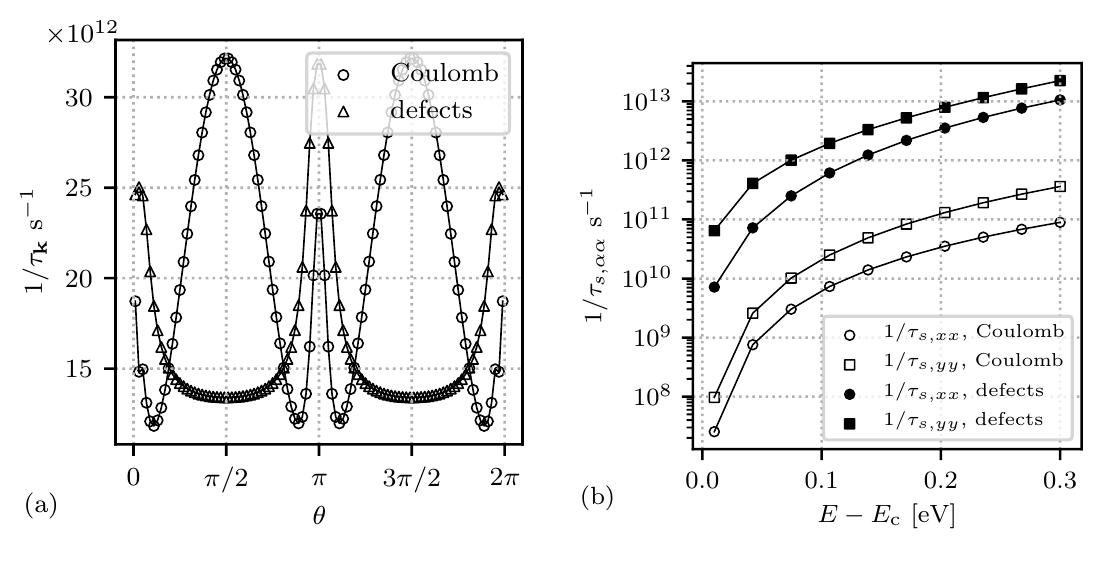} 
    \vspace{-0.1in}
    \caption[example] 
    {(a) Total $k$--dependent momentum scattering rate for monolayer Black phosphorus where $n_\text{i}=n_\text{d}=10^{10}$cm$^{-2}$.
    (b) Spin relaxation rate for conduction band of monolayer Black Phosphorus with effective masses $m_x=1.26m$ and $m_y=0.17m$. 
    The horizontal axis represents energy level relative to the band edge.}
    \label{fig:bp} 
\vspace{-0.1in}
\end{figure*}

We apply the calculations to study spin relaxation in monolayer BP due to the DP mechanism.
We only consider the conduction band of monolayer BP with effective mass of electrons $m_x=1.26m$ and $m_y=0.17m$, where $m$ is the free electron mass.\cite{popovic2015electronic} 
We assume that the monolayer is deposited on an hBN substrate\cite{avsar2017gate} with relative permittivity of $\epsilon_\text{r}=3.8$.
First, we plot the total momentum scattering rate given as $1/\tau_{\vb*{k}}=\sum_{\vb*{k'}\not=\vb*{k}}W_{\vb*{kk'}}$ in Fig. \ref{fig:bp}a. 
Here, we use typical values of $n_\text{i}=n_\text{d}=10^{10}$ cm$^{-2}$, $V_0=10$ eV,\cite{yuan2015transport, hwang2008single} $d=\sigma=10a_0$ ($a_0$ is the Bohr radius), and $E - E_\text{c}=0.1$ eV.
As seen from the figure, the momentum scattering rate $1/\tau_{\vb*{k}}$ shows a high anisotropy which consequently affects the spin relaxation.
Next, we plot spin relaxation rates for initial polarization along the $\vu*{x}$ and $\vu*{y}$ axes i.e. $1/\tau_{s,xx}$ and $1/\tau_{s,yy}$. 
Figure \ref{fig:bp}b depicts the energy dependence of spin relaxation rate which is proportional to $\lambda^2_\text{R}\epsilon^2_\text{r}/n_\text{i}$ for the Coulomb potential and $\lambda^2_\text{R}/n_\text{d}$ for defects. 
The inverse proportionality of spin relaxation rate to  $n_\text{i}$ and $n_\text{d}$ is the signature of the DP mechanism. 
The horizontal axis represents energy level relative to the conduction band edge, i.e. $E - E_\text{c}$.
We can see from the figure that the spin relaxation rate is highly dependent on the energy level. 
Increasing the energy level by $0.3$ eV, raises the spin relaxation rate by few orders of magnitude depending on the scattering potential. 
These results can also describe the spin relaxation in few layers BP whose band structure is also elliptic with similar anisotropy to that of monolayer BP but with different band gap which is dependent on the number of layers.\cite{qiao2014high, tran2014layer}
We note that as the energy changes, the ratio of in--plane spin relaxation times, $\tau_{s, yy}/\tau_{s, xx}$, remains constant.
For an electric field of $1$ V/nm, the Rashba strength of $\hbar\lambda_\text{R}\sim1$ meV$\cdot$\AA~can be achieved.\cite{kurpas2018spin}
For typical values of $n_\text{i}=n_\text{d}=10^{10}$cm$^{-2}$, $V_0=10$ eV, and $E - E_\text{c}=0.1$ eV the spin relaxation rates  are $1/\tau_{s,xx}\simeq 7\times10^{9}$ s$^{-1}$ and $1/\tau_{s,yy}\simeq 3\times10^{10}$ s$^{-1}$ for the Coulomb potential and $1/\tau_{s,xx}\simeq 6\times10^{11}$ s$^{-1}$ and $1/\tau_{s,yy}\simeq 2\times10^{12}$ s$^{-1}$ for short--range defects where $\sigma=10a_0$.
The corresponding values of momentum scattering $1/\overline{\tau_{\vb*{k}}}$ are on the order of $10^{13}$ s$^{-1}$ which validates our assumption of strong momentum scattering.

Mechanical strain can alter the band gap and the effective carrier mass in monolayer BP. 
First principle calculations \cite{peng2014strain} have shown that the band gap decreases with increasing strain (both tensile and compressive) on the lattice.
However, the effective masses undergo sharp non--monotonic transitions at certain values of strain. 
Once the effect of strain on the effective masses is determined, we can find the corresponding effect on the spin relaxation.
For example, according to Ref. \onlinecite{peng2014strain}, an $8$\% tensile strain along the $\vu*{x}$ axis (zigzag direction) would change the effective masses $m_x=1.26m$ and $m_y=0.17m$ to considerably different values $m'_x=0.2m$ and $m'_y=1.05m$. 
Therefore, for a Coulomb dominated monolayer BP under strain we obtain $\tau'_{s,yy}/\tau'_{s,xx}=(m'_y/m'_x)^{0.5}=2.3$.

\section{conclusion}
In conclusion, spin dynamics in a 2D elliptic band structure, such as few--layer Black Phosphorus (BP), is studied due to the D'yakonov--Perel' mechanism. 
The elliptic band structure is characterized with in--plane effective masses $m_x$ and $m_y$. 
Two different scattering potentials namely the Coulomb potential and neutral defects are incorporated in the calculations.
Representing spin polarized ensemble with density matrices and using the time evolution equation of the ensemble, spin relaxation time $\tau_{s,\alpha\alpha}$ is calculated for an ensemble initially polarized along $\vu*{\alpha}$ axis. 
Spin relaxation is shown to be slower in the direction of the heavier effective mass.
More specifically, the in--plane anisotropy in spin relaxation time $\tau_{s,yy}/\tau_{s,xx}$ scales proportional to $(m_y/m_x)^\nu$ where $\nu$ depends on the scattering potential, i.e. $\nu\simeq0.5$ for the Coulomb potential and $\nu\simeq0.5-1.5$ for defects with different ranges. 
Effects of spin dephasing are not considered implying that the off--diagonal elements of the
{spin relaxation rate} are considered zero, i.e. $1/\tau_{s,\alpha\beta}=0$ for $\alpha\bot\beta$.
For the isotropic case, $m_y/m_x=1$, the well known result $\tau_{s,xx}=\tau_{s,yy}=2\tau_{s,zz}$ is reproduced. 
More generally, a spin ensemble initially polarized along the $\vu*{z}$ axis relaxes faster than any other directions because $1/\tau_{s,zz} = 1/\tau_{s,xx} + 1/\tau_{s,yy}$.
These calculations are applied to study spin relaxation in monolayer BP. For typical values of Rashba spin--orbit coupling, $\hbar\lambda_R=1$ meV$\cdot$\AA, and charged impurity concentration equal to $10^{10}$ cm$^{-2}$, we obtain $1/\tau_{s,xx}=7\times 10^{9}$ s$^{-1}$ and $1/\tau_{s,yy}=3\times 10^{10}$ s$^{-1}$. These numbers are comparable in magnitude to those predicted from the Elliott-Yafet mechanism in BP.  
Our results can be readily used to study the effect of strain on the spin relaxation anisotropy provided that the effective masses $m_x$ and $m_y$ are known as functions of strain. 
These results give insight in engineering spin transport media using few--layer BP and other similar 2D semiconductors with elliptic anisotropy.

\begin{acknowledgments}
The authors acknowledge the funding support from the MRSEC Program of the National Science Foundation under Award Number DMR-1420073.
\end{acknowledgments}

\appendix*
\section{Anisotropic Rashba Spin--Orbit Coupling}
The Hamiltonian for a two--dimensional crystal with lattice potential $V_0(\vb*{r})$ including Rashba spin--orbit coupling can be written as
\begin{equation}
     \mathcal{H} = \frac{p^2}{2m_0} + V_0(\vb*{r}) +  \frac{\lambda_\text{R}}{2}(\vb*{p}\times \vu*{z})\vdot\vb*{\sigma}\,,
\end{equation}
where $\lambda_\text{R}$ is the strength of Rashba spin--orbit term and depends on both $V_0(\vb*{r})$ and the external electric field. 
Applying the Hamiltonian on the Bloch wave functions $\psi(\vb*{r})=e^{i\vb*{k}\vdot\vb*{r}}u_{\vb*{k}}(\vb*{r})$, we obtain the Schr\"{o}dinger's equation for the lattice periodic functions $u_{\vb*{k}}(\vb*{r})=\ip{\vb*{r}}{n,\vb*{k}}$, i.e. $H\ket{n, \vb*{k}} = E_n(\vb*{k})\ket{n,\vb*{k}}$, where 
\begin{equation}
\label{eq:hamiltonian}
    H = \underbrace{\frac{p^2}{2m_0} + V_0(\vb*{r})}_{H_0} + \frac{\hbar^2k^2}{2m_0} + 
    \underbrace{\frac{\hbar}{m_0}\vb*{k}\vdot\vb*{p}}_{H_{\vb*{k}\vdot\vb*{p}}} + \underbrace{\frac{\lambda_\text{R}}{2}(\vb*{p}\times \vu*{z})\vdot\vb*{\sigma}}_{H_\text{R}}\cdot
\end{equation}
Generally, for light atoms like phosphorus, we can assume that $H_{\vb*{k}\vdot\vb*{p}}\gg H_\text{R}$. Therefore, in the absence of spin--orbit coupling, the eigenvalues and eigenkets of Hamiltonian \ref{eq:hamiltonian} are given in terms of the solutions to $H_0$ by using the $\vb*{k}\vdot\vb*{p}$ perturbation theory. The band structure of the $n^\text{th}$ band about $\vb*{k} = \vb*{0}$ is given as 
\begin{equation}
\begin{split}
    E_n(\vb*{k}) = & E_n(\vb*{0}) + \frac{\hbar^2k^2}{2m_0} + 
    \frac{\hbar}{m_0}\vb*{k}\vdot\mel{n, \vb*{0}}{\vb*{p}}{n, \vb*{0}} \\
    & + \frac{\hbar^2}{m_0^2}\sum_{n'\not=n}\frac{\qty|\vb*{k}\vdot\mel{n,\vb*{0}}{\vb*{p}}{n',\vb*{0}}|^2}{E_n(\vb*{0}) - E_{n'}(\vb*{0})}\cdot
\end{split}
\end{equation}
Assuming that the point $\vb*{k}=\vb*{0}$ is an extremum,  the first order term vanishes. Therefore, to the leading order in $\vb*{k}$ we obtain 
\begin{equation}
    E_n(\vb*{k}) = E_n(\vb*{0}) + \sum_{i,j}  \frac{\hbar^2k_{i}k_{j}}{2m_{n,ij}}\,,
\end{equation}
where the $m_{n,ij}$ parameters are the elements of the effective mass tensor given as
\begin{equation}
    \frac{1}{m_{n,ij}} = \frac{\delta_{ij}}{m_0} + \frac{2}{m_0^2}\sum_{n'\not=n}\frac{\mel{n,\vb*{0}}{p_i}{n',\vb*{0}}\mel{n',\vb*{0}}{p_j}{n,\vb*{0}}}{E_n(\vb*{0}) - E_{n'}(\vb*{0})}\cdot
\end{equation}
The eigenkets to the leading order in $\vb*{k}$ are given as 
\begin{equation}
\label{eq:ket}
\begin{split}
    \ket{n,\vb*{k}} = \frac{1}{\sqrt{N}}\qty(\ket{n, \vb*{0}} + \frac{\hbar}{m_0}\sum_{n'\not=n}\ket{n',\vb*{0}}\frac{ \vb*{k}\vdot\mel{n',\vb*{0}}{\vb*{p}}{n,\vb*{0}}}{E_n(\vb*{0}) - E_{n'}(\vb*{0})})\,, 
\end{split}
\end{equation}
where $N$ is the normalization factor.
In the absence of spin--orbit coupling, each band is doubly degenerate. Therefore, the spin--dependent eigenkets are $\ket{n,\vb*{k}}\otimes\ket{\pm}$. 
Representing $H_\text{R}$ in the $\ket{n,\vb*{k}}\otimes\ket{\pm}$ basis, we obtain
\begin{equation}
\begin{split}
    \widetilde{H}_\text{R} = \frac{\lambda_\text{R}}{2} &
    \big((\hbar k_y + \ev{p_y}_{n,\vb*{k}})\sigma_x
    - (\hbar k_x + \ev{p_x}_{n,\vb*{k}})\sigma_y\big)
\end{split}\cdot
\end{equation}
The expectation values $\ev{p_i}_{n,\vb*{k}}$ are calculated using Eq. \ref{eq:ket} to the leading order in $\vb*{k}$ as follows
\begin{subequations}
\begin{equation}
    \ev{p_x}_{n,\vb*{k}} = \frac{\hbar k_x}{m_0}\sum_{m\not=n}\frac{|\mel{m,\vb*{0}}{p_x}{n, \vb*{0}}|^2}{E_{n,\vb*{0}} - E_{m,\vb*{0}}} + \frac{\hbar k_y}{m_0}\frac{m_0^2}{2m_{yx}}\,,
\end{equation}
\begin{equation}
    \ev{p_y}_{n,\vb*{k}} = \frac{\hbar k_x}{m_0}\frac{m_0^2}{2m_{xy}} + \frac{\hbar k_y}{m_0}\sum_{m\not=n}\frac{|\mel{m,\vb*{0}}{p_y}{n, \vb*{0}}|^2}{E_{n,\vb*{0}} - E_{m,\vb*{0}}}\cdot
\end{equation}
\end{subequations}
Therefore, 
\begin{equation}
\label{eq:soc}
\begin{split}
    \widetilde{H}_\text{R} = \frac{\lambda_\text{R}}{2}
    & \bigg((\frac{m_0}{m_y})k_y\sigma_x
    - (\frac{m_0}{m_x})k_x\sigma_y \\
    + & (\frac{m_0}{2m_{xy}})k_x\sigma_x
    + (\frac{m_0}{2m_{yx}})k_y\sigma_y\bigg)
\end{split}\cdot
\end{equation}
Provided that the $x$-- and $y$--directions represent the principal axes, the off--diagonal elements of the effective mass tensor vanish, i.e. $m_{xy}=m_{yx}=0$. Finally, the $k$--space Hamiltonian of the anisotropic system is 
\begin{equation}
\begin{split}
    H = & \underbrace{E_n(\vb*{0}) + \frac{\hbar^2k_x^2}{2m_x} + \frac{\hbar^2k_y^2}{2m_y}}_{H_0} \\ + & \underbrace{\frac{\lambda_\text{R}}{2}\qty((\frac{m_0}{m_y})k_y\sigma_x
    - (\frac{m_0}{m_x})k_x\sigma_y)}_{H_{\vb*{k}}}\cdot
\end{split}
\end{equation}

\bibliographystyle{apsrev4-1}
\bibliography{main}

\begin{thebibliography}{28}%
\makeatletter
\providecommand \@ifxundefined [1]{%
 \@ifx{#1\undefined}
}%
\providecommand \@ifnum [1]{%
 \ifnum #1\expandafter \@firstoftwo
 \else \expandafter \@secondoftwo
 \fi
}%
\providecommand \@ifx [1]{%
 \ifx #1\expandafter \@firstoftwo
 \else \expandafter \@secondoftwo
 \fi
}%
\providecommand \natexlab [1]{#1}%
\providecommand \enquote  [1]{``#1''}%
\providecommand \bibnamefont  [1]{#1}%
\providecommand \bibfnamefont [1]{#1}%
\providecommand \citenamefont [1]{#1}%
\providecommand \href@noop [0]{\@secondoftwo}%
\providecommand \href [0]{\begingroup \@sanitize@url \@href}%
\providecommand \@href[1]{\@@startlink{#1}\@@href}%
\providecommand \@@href[1]{\endgroup#1\@@endlink}%
\providecommand \@sanitize@url [0]{\catcode `\\12\catcode `\$12\catcode
  `\&12\catcode `\#12\catcode `\^12\catcode `\_12\catcode `\%12\relax}%
\providecommand \@@startlink[1]{}%
\providecommand \@@endlink[0]{}%
\providecommand \url  [0]{\begingroup\@sanitize@url \@url }%
\providecommand \@url [1]{\endgroup\@href {#1}{\urlprefix }}%
\providecommand \urlprefix  [0]{URL }%
\providecommand \Eprint [0]{\href }%
\providecommand \doibase [0]{http://dx.doi.org/}%
\providecommand \selectlanguage [0]{\@gobble}%
\providecommand \bibinfo  [0]{\@secondoftwo}%
\providecommand \bibfield  [0]{\@secondoftwo}%
\providecommand \translation [1]{[#1]}%
\providecommand \BibitemOpen [0]{}%
\providecommand \bibitemStop [0]{}%
\providecommand \bibitemNoStop [0]{.\EOS\space}%
\providecommand \EOS [0]{\spacefactor3000\relax}%
\providecommand \BibitemShut  [1]{\csname bibitem#1\endcsname}%
\let\auto@bib@innerbib\@empty
\bibitem [{\citenamefont {Nishikawa}\ \emph {et~al.}(1995)\citenamefont
  {Nishikawa}, \citenamefont {Tackeuchi}, \citenamefont {Nakamura},
  \citenamefont {Muto},\ and\ \citenamefont {Yokoyama}}]{nishikawa1995all}%
  \BibitemOpen
  \bibfield  {author} {\bibinfo {author} {\bibfnamefont {Y.}~\bibnamefont
  {Nishikawa}}, \bibinfo {author} {\bibfnamefont {A.}~\bibnamefont
  {Tackeuchi}}, \bibinfo {author} {\bibfnamefont {S.}~\bibnamefont {Nakamura}},
  \bibinfo {author} {\bibfnamefont {S.}~\bibnamefont {Muto}}, \ and\ \bibinfo
  {author} {\bibfnamefont {N.}~\bibnamefont {Yokoyama}},\ }\href@noop {}
  {\bibfield  {journal} {\bibinfo  {journal} {Applied physics letters}\
  }\textbf {\bibinfo {volume} {66}},\ \bibinfo {pages} {839} (\bibinfo {year}
  {1995})}\BibitemShut {NoStop}%
\bibitem [{\citenamefont {Hall}\ \emph {et~al.}(1999)\citenamefont {Hall},
  \citenamefont {Leonard}, \citenamefont {van Driel}, \citenamefont {Kost},
  \citenamefont {Selvig},\ and\ \citenamefont {Chow}}]{hall1999subpicosecond}%
  \BibitemOpen
  \bibfield  {author} {\bibinfo {author} {\bibfnamefont {K.}~\bibnamefont
  {Hall}}, \bibinfo {author} {\bibfnamefont {S.}~\bibnamefont {Leonard}},
  \bibinfo {author} {\bibfnamefont {H.}~\bibnamefont {van Driel}}, \bibinfo
  {author} {\bibfnamefont {A.}~\bibnamefont {Kost}}, \bibinfo {author}
  {\bibfnamefont {E.}~\bibnamefont {Selvig}}, \ and\ \bibinfo {author}
  {\bibfnamefont {D.}~\bibnamefont {Chow}},\ }\href@noop {} {\bibfield
  {journal} {\bibinfo  {journal} {Applied Physics Letters}\ }\textbf {\bibinfo
  {volume} {75}},\ \bibinfo {pages} {4156} (\bibinfo {year}
  {1999})}\BibitemShut {NoStop}%
\bibitem [{\citenamefont {Min}\ \emph {et~al.}(2006)\citenamefont {Min},
  \citenamefont {Hill}, \citenamefont {Sinitsyn}, \citenamefont {Sahu},
  \citenamefont {Kleinman},\ and\ \citenamefont
  {MacDonald}}]{min2006intrinsic}%
  \BibitemOpen
  \bibfield  {author} {\bibinfo {author} {\bibfnamefont {H.}~\bibnamefont
  {Min}}, \bibinfo {author} {\bibfnamefont {J.}~\bibnamefont {Hill}}, \bibinfo
  {author} {\bibfnamefont {N.~A.}\ \bibnamefont {Sinitsyn}}, \bibinfo {author}
  {\bibfnamefont {B.}~\bibnamefont {Sahu}}, \bibinfo {author} {\bibfnamefont
  {L.}~\bibnamefont {Kleinman}}, \ and\ \bibinfo {author} {\bibfnamefont
  {A.~H.}\ \bibnamefont {MacDonald}},\ }\href@noop {} {\bibfield  {journal}
  {\bibinfo  {journal} {Physical Review B}\ }\textbf {\bibinfo {volume} {74}},\
  \bibinfo {pages} {165310} (\bibinfo {year} {2006})}\BibitemShut {NoStop}%
\bibitem [{\citenamefont {Han}\ and\ \citenamefont
  {Kawakami}(2011)}]{han2011spin}%
  \BibitemOpen
  \bibfield  {author} {\bibinfo {author} {\bibfnamefont {W.}~\bibnamefont
  {Han}}\ and\ \bibinfo {author} {\bibfnamefont {R.~K.}\ \bibnamefont
  {Kawakami}},\ }\href@noop {} {\bibfield  {journal} {\bibinfo  {journal}
  {Physical review letters}\ }\textbf {\bibinfo {volume} {107}},\ \bibinfo
  {pages} {047207} (\bibinfo {year} {2011})}\BibitemShut {NoStop}%
\bibitem [{\citenamefont {Avsar}\ \emph {et~al.}(2017)\citenamefont {Avsar},
  \citenamefont {Tan}, \citenamefont {Kurpas}, \citenamefont {Gmitra},
  \citenamefont {Watanabe}, \citenamefont {Taniguchi}, \citenamefont {Fabian},\
  and\ \citenamefont {{\"O}zyilmaz}}]{avsar2017gate}%
  \BibitemOpen
  \bibfield  {author} {\bibinfo {author} {\bibfnamefont {A.}~\bibnamefont
  {Avsar}}, \bibinfo {author} {\bibfnamefont {J.~Y.}\ \bibnamefont {Tan}},
  \bibinfo {author} {\bibfnamefont {M.}~\bibnamefont {Kurpas}}, \bibinfo
  {author} {\bibfnamefont {M.}~\bibnamefont {Gmitra}}, \bibinfo {author}
  {\bibfnamefont {K.}~\bibnamefont {Watanabe}}, \bibinfo {author}
  {\bibfnamefont {T.}~\bibnamefont {Taniguchi}}, \bibinfo {author}
  {\bibfnamefont {J.}~\bibnamefont {Fabian}}, \ and\ \bibinfo {author}
  {\bibfnamefont {B.}~\bibnamefont {{\"O}zyilmaz}},\ }\href@noop {} {\bibfield
  {journal} {\bibinfo  {journal} {Nature Physics}\ }\textbf {\bibinfo {volume}
  {13}},\ \bibinfo {pages} {nphys4141} (\bibinfo {year} {2017})}\BibitemShut
  {NoStop}%
\bibitem [{\citenamefont {Kurpas}\ \emph {et~al.}(2018)\citenamefont {Kurpas},
  \citenamefont {Fabian} \emph {et~al.}}]{kurpas2018spin}%
  \BibitemOpen
  \bibfield  {author} {\bibinfo {author} {\bibfnamefont {M.}~\bibnamefont
  {Kurpas}}, \bibinfo {author} {\bibfnamefont {J.}~\bibnamefont {Fabian}},
  \emph {et~al.},\ }\href@noop {} {\bibfield  {journal} {\bibinfo  {journal}
  {Journal of Physics D: Applied Physics}\ } (\bibinfo {year}
  {2018})}\BibitemShut {NoStop}%
\bibitem [{\citenamefont {Hill}\ \emph {et~al.}(2006)\citenamefont {Hill},
  \citenamefont {Geim}, \citenamefont {Novoselov}, \citenamefont {Schedin},\
  and\ \citenamefont {Blake}}]{hill2006graphene}%
  \BibitemOpen
  \bibfield  {author} {\bibinfo {author} {\bibfnamefont {E.~W.}\ \bibnamefont
  {Hill}}, \bibinfo {author} {\bibfnamefont {A.~K.}\ \bibnamefont {Geim}},
  \bibinfo {author} {\bibfnamefont {K.}~\bibnamefont {Novoselov}}, \bibinfo
  {author} {\bibfnamefont {F.}~\bibnamefont {Schedin}}, \ and\ \bibinfo
  {author} {\bibfnamefont {P.}~\bibnamefont {Blake}},\ }\href@noop {}
  {\bibfield  {journal} {\bibinfo  {journal} {IEEE Transactions on Magnetics}\
  }\textbf {\bibinfo {volume} {42}},\ \bibinfo {pages} {2694} (\bibinfo {year}
  {2006})}\BibitemShut {NoStop}%
\bibitem [{\citenamefont {Tombros}\ \emph {et~al.}(2007)\citenamefont
  {Tombros}, \citenamefont {Jozsa}, \citenamefont {Popinciuc}, \citenamefont
  {Jonkman},\ and\ \citenamefont {Van~Wees}}]{tombros2007electronic}%
  \BibitemOpen
  \bibfield  {author} {\bibinfo {author} {\bibfnamefont {N.}~\bibnamefont
  {Tombros}}, \bibinfo {author} {\bibfnamefont {C.}~\bibnamefont {Jozsa}},
  \bibinfo {author} {\bibfnamefont {M.}~\bibnamefont {Popinciuc}}, \bibinfo
  {author} {\bibfnamefont {H.~T.}\ \bibnamefont {Jonkman}}, \ and\ \bibinfo
  {author} {\bibfnamefont {B.~J.}\ \bibnamefont {Van~Wees}},\ }\href@noop {}
  {\bibfield  {journal} {\bibinfo  {journal} {Nature}\ }\textbf {\bibinfo
  {volume} {448}},\ \bibinfo {pages} {571} (\bibinfo {year}
  {2007})}\BibitemShut {NoStop}%
\bibitem [{\citenamefont {Ingla-Ayn{\'e}s}\ \emph {et~al.}(2015)\citenamefont
  {Ingla-Ayn{\'e}s}, \citenamefont {Guimar{\~a}es}, \citenamefont {Meijerink},
  \citenamefont {Zomer},\ and\ \citenamefont {van Wees}}]{ingla201524}%
  \BibitemOpen
  \bibfield  {author} {\bibinfo {author} {\bibfnamefont {J.}~\bibnamefont
  {Ingla-Ayn{\'e}s}}, \bibinfo {author} {\bibfnamefont {M.~H.}\ \bibnamefont
  {Guimar{\~a}es}}, \bibinfo {author} {\bibfnamefont {R.~J.}\ \bibnamefont
  {Meijerink}}, \bibinfo {author} {\bibfnamefont {P.~J.}\ \bibnamefont
  {Zomer}}, \ and\ \bibinfo {author} {\bibfnamefont {B.~J.}\ \bibnamefont {van
  Wees}},\ }\href@noop {} {\bibfield  {journal} {\bibinfo  {journal} {Physical
  Review B}\ }\textbf {\bibinfo {volume} {92}},\ \bibinfo {pages} {201410}
  (\bibinfo {year} {2015})}\BibitemShut {NoStop}%
\bibitem [{\citenamefont {Dr\"ogeler}\ \emph {et~al.}(2016)\citenamefont
  {Dr\"ogeler}, \citenamefont {Franzen}, \citenamefont {Volmer}, \citenamefont
  {Pohlmann}, \citenamefont {Banszerus}, \citenamefont {Wolter}, \citenamefont
  {Watanabe}, \citenamefont {Taniguchi}, \citenamefont {Stampfer},\ and\
  \citenamefont {Beschoten}}]{drogeler2016spin}%
  \BibitemOpen
  \bibfield  {author} {\bibinfo {author} {\bibfnamefont {M.}~\bibnamefont
  {Dr\"ogeler}}, \bibinfo {author} {\bibfnamefont {C.}~\bibnamefont {Franzen}},
  \bibinfo {author} {\bibfnamefont {F.}~\bibnamefont {Volmer}}, \bibinfo
  {author} {\bibfnamefont {T.}~\bibnamefont {Pohlmann}}, \bibinfo {author}
  {\bibfnamefont {L.}~\bibnamefont {Banszerus}}, \bibinfo {author}
  {\bibfnamefont {M.}~\bibnamefont {Wolter}}, \bibinfo {author} {\bibfnamefont
  {K.}~\bibnamefont {Watanabe}}, \bibinfo {author} {\bibfnamefont
  {T.}~\bibnamefont {Taniguchi}}, \bibinfo {author} {\bibfnamefont
  {C.}~\bibnamefont {Stampfer}}, \ and\ \bibinfo {author} {\bibfnamefont
  {B.}~\bibnamefont {Beschoten}},\ }\href@noop {} {\bibfield  {journal}
  {\bibinfo  {journal} {Nano letters}\ }\textbf {\bibinfo {volume} {16}},\
  \bibinfo {pages} {3533} (\bibinfo {year} {2016})}\BibitemShut {NoStop}%
\bibitem [{\citenamefont {Peng}\ \emph {et~al.}(2014)\citenamefont {Peng},
  \citenamefont {Wei},\ and\ \citenamefont {Copple}}]{peng2014strain}%
  \BibitemOpen
  \bibfield  {author} {\bibinfo {author} {\bibfnamefont {X.}~\bibnamefont
  {Peng}}, \bibinfo {author} {\bibfnamefont {Q.}~\bibnamefont {Wei}}, \ and\
  \bibinfo {author} {\bibfnamefont {A.}~\bibnamefont {Copple}},\ }\href@noop {}
  {\bibfield  {journal} {\bibinfo  {journal} {Physical Review B}\ }\textbf
  {\bibinfo {volume} {90}},\ \bibinfo {pages} {085402} (\bibinfo {year}
  {2014})}\BibitemShut {NoStop}%
\bibitem [{\citenamefont {Ong}\ \emph {et~al.}(2014)\citenamefont {Ong},
  \citenamefont {Zhang},\ and\ \citenamefont {Zhang}}]{ong2014anisotropic}%
  \BibitemOpen
  \bibfield  {author} {\bibinfo {author} {\bibfnamefont {Z.-Y.}\ \bibnamefont
  {Ong}}, \bibinfo {author} {\bibfnamefont {G.}~\bibnamefont {Zhang}}, \ and\
  \bibinfo {author} {\bibfnamefont {Y.~W.}\ \bibnamefont {Zhang}},\ }\href@noop
  {} {\bibfield  {journal} {\bibinfo  {journal} {Journal of Applied Physics}\
  }\textbf {\bibinfo {volume} {116}},\ \bibinfo {pages} {214505} (\bibinfo
  {year} {2014})}\BibitemShut {NoStop}%
\bibitem [{\citenamefont {Liu}\ \emph {et~al.}(2016)\citenamefont {Liu},
  \citenamefont {Low},\ and\ \citenamefont {Ruden}}]{liu2016mobility}%
  \BibitemOpen
  \bibfield  {author} {\bibinfo {author} {\bibfnamefont {Y.}~\bibnamefont
  {Liu}}, \bibinfo {author} {\bibfnamefont {T.}~\bibnamefont {Low}}, \ and\
  \bibinfo {author} {\bibfnamefont {P.~P.}\ \bibnamefont {Ruden}},\ }\href@noop
  {} {\bibfield  {journal} {\bibinfo  {journal} {Physical Review B}\ }\textbf
  {\bibinfo {volume} {93}},\ \bibinfo {pages} {165402} (\bibinfo {year}
  {2016})}\BibitemShut {NoStop}%
\bibitem [{\citenamefont {Liu}\ and\ \citenamefont
  {Ruden}(2017)}]{liu2017temperature}%
  \BibitemOpen
  \bibfield  {author} {\bibinfo {author} {\bibfnamefont {Y.}~\bibnamefont
  {Liu}}\ and\ \bibinfo {author} {\bibfnamefont {P.~P.}\ \bibnamefont
  {Ruden}},\ }\href@noop {} {\bibfield  {journal} {\bibinfo  {journal}
  {Physical Review B}\ }\textbf {\bibinfo {volume} {95}},\ \bibinfo {pages}
  {165446} (\bibinfo {year} {2017})}\BibitemShut {NoStop}%
\bibitem [{\citenamefont {Popovi{\'c}}\ \emph {et~al.}(2015)\citenamefont
  {Popovi{\'c}}, \citenamefont {Kurdestany},\ and\ \citenamefont
  {Satpathy}}]{popovic2015electronic}%
  \BibitemOpen
  \bibfield  {author} {\bibinfo {author} {\bibfnamefont {Z.}~\bibnamefont
  {Popovi{\'c}}}, \bibinfo {author} {\bibfnamefont {J.~M.}\ \bibnamefont
  {Kurdestany}}, \ and\ \bibinfo {author} {\bibfnamefont {S.}~\bibnamefont
  {Satpathy}},\ }\href@noop {} {\bibfield  {journal} {\bibinfo  {journal}
  {Physical Review B}\ }\textbf {\bibinfo {volume} {92}},\ \bibinfo {pages}
  {035135} (\bibinfo {year} {2015})}\BibitemShut {NoStop}%
\bibitem [{\citenamefont {Kurpas}\ \emph {et~al.}(2016)\citenamefont {Kurpas},
  \citenamefont {Gmitra},\ and\ \citenamefont {Fabian}}]{kurpas2016spin}%
  \BibitemOpen
  \bibfield  {author} {\bibinfo {author} {\bibfnamefont {M.}~\bibnamefont
  {Kurpas}}, \bibinfo {author} {\bibfnamefont {M.}~\bibnamefont {Gmitra}}, \
  and\ \bibinfo {author} {\bibfnamefont {J.}~\bibnamefont {Fabian}},\
  }\href@noop {} {\bibfield  {journal} {\bibinfo  {journal} {Physical Review
  B}\ }\textbf {\bibinfo {volume} {94}},\ \bibinfo {pages} {155423} (\bibinfo
  {year} {2016})}\BibitemShut {NoStop}%
\bibitem [{\citenamefont {Dyakonov}\ and\ \citenamefont
  {Perel}(1972)}]{dyakonov1972spin}%
  \BibitemOpen
  \bibfield  {author} {\bibinfo {author} {\bibfnamefont {M.}~\bibnamefont
  {Dyakonov}}\ and\ \bibinfo {author} {\bibfnamefont {V.}~\bibnamefont
  {Perel}},\ }\href@noop {} {\bibfield  {journal} {\bibinfo  {journal} {Soviet
  Physics Solid State, USSR}\ }\textbf {\bibinfo {volume} {13}},\ \bibinfo
  {pages} {3023} (\bibinfo {year} {1972})}\BibitemShut {NoStop}%
\bibitem [{\citenamefont {Bychkov}\ and\ \citenamefont
  {Rashba}(1984)}]{bychkov1984properties}%
  \BibitemOpen
  \bibfield  {author} {\bibinfo {author} {\bibfnamefont {Y.~A.}\ \bibnamefont
  {Bychkov}}\ and\ \bibinfo {author} {\bibfnamefont {E.}~\bibnamefont
  {Rashba}},\ }\href@noop {} {\bibfield  {journal} {\bibinfo  {journal} {JETP
  lett}\ }\textbf {\bibinfo {volume} {39}},\ \bibinfo {pages} {78} (\bibinfo
  {year} {1984})}\BibitemShut {NoStop}%
\bibitem [{\citenamefont {Elliott}(1954)}]{elliott1954theory}%
  \BibitemOpen
  \bibfield  {author} {\bibinfo {author} {\bibfnamefont {R.~J.}\ \bibnamefont
  {Elliott}},\ }\href@noop {} {\bibfield  {journal} {\bibinfo  {journal}
  {Physical Review}\ }\textbf {\bibinfo {volume} {96}},\ \bibinfo {pages} {266}
  (\bibinfo {year} {1954})}\BibitemShut {NoStop}%
\bibitem [{\citenamefont {Li}\ and\ \citenamefont
  {Appelbaum}(2014)}]{li2014electrons}%
  \BibitemOpen
  \bibfield  {author} {\bibinfo {author} {\bibfnamefont {P.}~\bibnamefont
  {Li}}\ and\ \bibinfo {author} {\bibfnamefont {I.}~\bibnamefont {Appelbaum}},\
  }\href@noop {} {\bibfield  {journal} {\bibinfo  {journal} {Physical Review
  B}\ }\textbf {\bibinfo {volume} {90}},\ \bibinfo {pages} {115439} (\bibinfo
  {year} {2014})}\BibitemShut {NoStop}%
\bibitem [{\citenamefont {Fabian}\ \emph {et~al.}(2007)\citenamefont {Fabian},
  \citenamefont {Matos-Abiague}, \citenamefont {Ertler}, \citenamefont
  {Stano},\ and\ \citenamefont {Zutic}}]{fabian2007semiconductor}%
  \BibitemOpen
  \bibfield  {author} {\bibinfo {author} {\bibfnamefont {J.}~\bibnamefont
  {Fabian}}, \bibinfo {author} {\bibfnamefont {A.}~\bibnamefont
  {Matos-Abiague}}, \bibinfo {author} {\bibfnamefont {C.}~\bibnamefont
  {Ertler}}, \bibinfo {author} {\bibfnamefont {P.}~\bibnamefont {Stano}}, \
  and\ \bibinfo {author} {\bibfnamefont {I.}~\bibnamefont {Zutic}},\
  }\href@noop {} {\bibfield  {journal} {\bibinfo  {journal} {Acta Physica
  Slovaca}\ }\textbf {\bibinfo {volume} {57}},\ \bibinfo {pages} {565}
  (\bibinfo {year} {2007})}\BibitemShut {NoStop}%
\bibitem [{\citenamefont {Averkiev}\ \emph {et~al.}(2002)\citenamefont
  {Averkiev}, \citenamefont {Golub},\ and\ \citenamefont
  {Willander}}]{averkiev2002spin}%
  \BibitemOpen
  \bibfield  {author} {\bibinfo {author} {\bibfnamefont {N.}~\bibnamefont
  {Averkiev}}, \bibinfo {author} {\bibfnamefont {L.}~\bibnamefont {Golub}}, \
  and\ \bibinfo {author} {\bibfnamefont {M.}~\bibnamefont {Willander}},\
  }\href@noop {} {\bibfield  {journal} {\bibinfo  {journal} {Journal of
  physics: condensed matter}\ }\textbf {\bibinfo {volume} {14}},\ \bibinfo
  {pages} {R271} (\bibinfo {year} {2002})}\BibitemShut {NoStop}%
\bibitem [{\citenamefont {Seixas}\ \emph {et~al.}(2015)\citenamefont {Seixas},
  \citenamefont {Carvalho},\ and\ \citenamefont {Neto}}]{seixas2015atomically}%
  \BibitemOpen
  \bibfield  {author} {\bibinfo {author} {\bibfnamefont {L.}~\bibnamefont
  {Seixas}}, \bibinfo {author} {\bibfnamefont {A.}~\bibnamefont {Carvalho}}, \
  and\ \bibinfo {author} {\bibfnamefont {A.~C.}\ \bibnamefont {Neto}},\
  }\href@noop {} {\bibfield  {journal} {\bibinfo  {journal} {Physical Review
  B}\ }\textbf {\bibinfo {volume} {91}},\ \bibinfo {pages} {155138} (\bibinfo
  {year} {2015})}\BibitemShut {NoStop}%
\bibitem [{\citenamefont {Ando}\ \emph {et~al.}(1982)\citenamefont {Ando},
  \citenamefont {Fowler},\ and\ \citenamefont {Stern}}]{ando1982electronic}%
  \BibitemOpen
  \bibfield  {author} {\bibinfo {author} {\bibfnamefont {T.}~\bibnamefont
  {Ando}}, \bibinfo {author} {\bibfnamefont {A.~B.}\ \bibnamefont {Fowler}}, \
  and\ \bibinfo {author} {\bibfnamefont {F.}~\bibnamefont {Stern}},\
  }\href@noop {} {\bibfield  {journal} {\bibinfo  {journal} {Reviews of Modern
  Physics}\ }\textbf {\bibinfo {volume} {54}},\ \bibinfo {pages} {437}
  (\bibinfo {year} {1982})}\BibitemShut {NoStop}%
\bibitem [{\citenamefont {Yuan}\ \emph {et~al.}(2015)\citenamefont {Yuan},
  \citenamefont {Rudenko},\ and\ \citenamefont
  {Katsnelson}}]{yuan2015transport}%
  \BibitemOpen
  \bibfield  {author} {\bibinfo {author} {\bibfnamefont {S.}~\bibnamefont
  {Yuan}}, \bibinfo {author} {\bibfnamefont {A.}~\bibnamefont {Rudenko}}, \
  and\ \bibinfo {author} {\bibfnamefont {M.}~\bibnamefont {Katsnelson}},\
  }\href@noop {} {\bibfield  {journal} {\bibinfo  {journal} {Physical Review
  B}\ }\textbf {\bibinfo {volume} {91}},\ \bibinfo {pages} {115436} (\bibinfo
  {year} {2015})}\BibitemShut {NoStop}%
\bibitem [{\citenamefont {Hwang}\ and\ \citenamefont
  {Sarma}(2008)}]{hwang2008single}%
  \BibitemOpen
  \bibfield  {author} {\bibinfo {author} {\bibfnamefont {E.}~\bibnamefont
  {Hwang}}\ and\ \bibinfo {author} {\bibfnamefont {S.~D.}\ \bibnamefont
  {Sarma}},\ }\href@noop {} {\bibfield  {journal} {\bibinfo  {journal}
  {Physical Review B}\ }\textbf {\bibinfo {volume} {77}},\ \bibinfo {pages}
  {195412} (\bibinfo {year} {2008})}\BibitemShut {NoStop}%
\bibitem [{\citenamefont {Qiao}\ \emph {et~al.}(2014)\citenamefont {Qiao},
  \citenamefont {Kong}, \citenamefont {Hu}, \citenamefont {Yang},\ and\
  \citenamefont {Ji}}]{qiao2014high}%
  \BibitemOpen
  \bibfield  {author} {\bibinfo {author} {\bibfnamefont {J.}~\bibnamefont
  {Qiao}}, \bibinfo {author} {\bibfnamefont {X.}~\bibnamefont {Kong}}, \bibinfo
  {author} {\bibfnamefont {Z.-X.}\ \bibnamefont {Hu}}, \bibinfo {author}
  {\bibfnamefont {F.}~\bibnamefont {Yang}}, \ and\ \bibinfo {author}
  {\bibfnamefont {W.}~\bibnamefont {Ji}},\ }\href@noop {} {\bibfield  {journal}
  {\bibinfo  {journal} {Nature communications}\ }\textbf {\bibinfo {volume}
  {5}},\ \bibinfo {pages} {4475} (\bibinfo {year} {2014})}\BibitemShut
  {NoStop}%
\bibitem [{\citenamefont {Tran}\ \emph {et~al.}(2014)\citenamefont {Tran},
  \citenamefont {Soklaski}, \citenamefont {Liang},\ and\ \citenamefont
  {Yang}}]{tran2014layer}%
  \BibitemOpen
  \bibfield  {author} {\bibinfo {author} {\bibfnamefont {V.}~\bibnamefont
  {Tran}}, \bibinfo {author} {\bibfnamefont {R.}~\bibnamefont {Soklaski}},
  \bibinfo {author} {\bibfnamefont {Y.}~\bibnamefont {Liang}}, \ and\ \bibinfo
  {author} {\bibfnamefont {L.}~\bibnamefont {Yang}},\ }\href@noop {} {\bibfield
   {journal} {\bibinfo  {journal} {Physical Review B}\ }\textbf {\bibinfo
  {volume} {89}},\ \bibinfo {pages} {235319} (\bibinfo {year}
  {2014})}\BibitemShut {NoStop}%
\end{thebibliography}%
\nocite{*}

\end{document}